\pdfoutput=1

\documentclass[11pt,twoside,a4paper,cmspaper,final,collab]{cms-tdr}
\def\svnVersion{454502P}\def\cmsCernNoTag{CERN-EP-2017-080}\def\cmsCernDate{\today}\def\cmsMessage{Published in Physical Review Letters as \href{http://dx.doi.org/10.1103/PhysRevLett.120.142301}{\doi{10.1103/PhysRevLett.120.142301.}}}

\begin{document}\cmsNoteHeader{HIN-16-008}

\hyphenation{had-ron-i-za-tion}
\hyphenation{cal-or-i-me-ter}
\hyphenation{de-vices}
\RCS$HeadURL: svn+ssh://svn.cern.ch/reps/tdr2/papers/HIN-16-008/trunk/HIN-16-008.tex $
\RCS$Id: HIN-16-008.tex 445377 2018-02-11 22:09:05Z alverson $
\newlength\cmsFigWidth
\ifthenelse{\boolean{cms@external}}{\setlength\cmsFigWidth{0.85\columnwidth}}{\setlength\cmsFigWidth{0.4\textwidth}}
\ifthenelse{\boolean{cms@external}}{\providecommand{\cmsLeft}{top\xspace}}{\providecommand{\cmsLeft}{left\xspace}}
\ifthenelse{\boolean{cms@external}}{\providecommand{\cmsRight}{bottom\xspace}}{\providecommand{\cmsRight}{right\xspace}}

\cmsNoteHeader{HIN-16-008}

\newcommand{\sqrtsnn}{\ensuremath{\sqrt{\smash[b]{s_{_{\mathrm{NN}}}}}}\xspace}
\newcommand{\upsi}{\ensuremath{\varUpsilon}\xspace}
\newcommand{\upsin}{\ensuremath{\varUpsilon(\mathrm{nS})}\xspace}
\newcommand{\upsione}{\ensuremath{\varUpsilon(\mathrm{1S})}\xspace}
\newcommand{\upsitwo}{\ensuremath{\varUpsilon(\mathrm{2S})}\xspace}
\newcommand{\upsithree}{\ensuremath{\varUpsilon(\mathrm{3S})}\xspace}
\newcommand{\pp}{\ensuremath{\Pp\Pp}\xspace}
\newcommand{\PbPb}{\ensuremath{\mathrm{PbPb}}\xspace}
\newcommand{\Npart}{\ensuremath{N_{\text{part}}}\xspace}
\newcommand{\Tc}{\ensuremath{T_\mathrm{c}}\xspace}
\ifthenelse{\boolean{cms@external}}{\providecommand{\CL}{C.L.\xspace}}{\providecommand{\CL}{CL\xspace}}
\ifthenelse{\boolean{cms@external}}{\providecommand{\CLp}{C.L\xspace}}{\providecommand{\CLp}{CL\xspace}}

\title{
\texorpdfstring{Suppression of excited \upsi states relative to the ground state in \PbPb collisions at $\sqrtsnn= 5.02$\TeV}{Suppression of excited Upsilon states relative to the ground state in PbPb collisions at sqrt(sNN) = 5.02TeV}
}

\date{\today}

\abstract{The relative yields of \upsi mesons produced in \pp and \PbPb collisions at $\sqrtsnn = 5.02$\TeV and reconstructed via the dimuon decay channel are measured using data collected by the CMS experiment. Double ratios are formed by comparing the yields of the excited states, \upsitwo and \upsithree, to the ground state, \upsione, in both \PbPb and \pp collisions at the same center-of-mass energy. The double ratios, $[\upsin/\upsione]_{\PbPb}/[\upsin/\upsione]_{\pp}$, are measured to be $0.308 \pm 0.055\stat \pm  0.019\syst $ for the \upsitwo and less than 0.26 at 95\% confidence level for the \upsithree. No significant \upsithree signal is found in the \PbPb data. The double ratios are studied as a function of collision centrality, as well as \upsi transverse momentum and rapidity. No significant dependencies are observed.
}
\hypersetup{%
pdfauthor={CMS Collaboration},%
pdftitle={Suppression of excited Upsilon states relative to the ground state in PbPb collisions at sqrt(sNN) = 5.02 TeV},%
pdfsubject={CMS, bottomonium, Upsilons},%
pdfkeywords={CMS, physics, bottomonium, quarkonium suppression, quark-gluon plasma, heavy-ion collisions}}

\maketitle
A key expectation of quantum chromodynamics (QCD) is that at high temperature, $T$, the degrees of freedom
will change and color fields and forces can act over ranges greater than typical hadronic sizes, a phenomenon referred to as color deconfinement. Studies of relativistic heavy ion collisions are motivated in large part by the goal of developing a detailed understanding of the properties of the deconfined phase, the quark-gluon plasma (QGP). Heavy quarkonia are some of the most promising probes of deconfinement, and hence have been the focus of detailed scrutiny. Quarkonium production is studied because of its sensitivity to color deconfinement via QCD Debye screening, as first proposed in Ref.~\cite{Matsui:1986dk}. Most of the early studies have focused on the charmonium family, but the high energies and collision rates available at the LHC enable studies of bottomonium states ~\cite{Chatrchyan:2011pe,Chatrchyan:2013nza,Chatrchyan:2012lxa,Khachatryan:2016xxp,Abelev:2014nua}.
Measurements of bottomonium suppression were performed~\cite{Adamczyk:2013poh} also at RHIC, and will be continued with
upgraded detectors~\cite{Adare:2015kwa}. Comparisons of \upsi data at the different collision energies will help to elucidate the temperature dependence of the suppression effects.

A detailed study of the modification of quarkonia states from \pp to \PbPb collisions
can provide information about the onset and properties of the QGP~\cite{Brambilla:2010cs,Andronic:2015wma}. In particular, suppression of heavy quarkonia via QCD Debye screening, or any other modification of the heavy-quark potential, requires the presence of a color-deconfined phase. Furthermore, the specific level of suppression for a given state depends on the QGP temperature. It is expected that different states will dissociate at different temperatures, with a suppression pattern ordered sequentially with binding energy~\cite{Digal:2001ue,Mocsy:2007jz}. The sequential suppression pattern was first observed for the \upsin family by CMS~\cite{Chatrchyan:2012lxa,Khachatryan:2016xxp}.

Recent theoretical studies consider not only the screening effect on the real part of the heavy-quark potential, but also incorporate an imaginary part~\cite{Laine:2006ns,Brambilla:2008cx,Dumitru:2009fy,Rothkopf:2011db,Burnier:2015tda}, which represents effects such as Landau damping and gluodissociation of the quarkonium states. These mechanisms broaden the width of the states and
also contribute to the suppression of the observed yields. A recent calculation~\cite{Burnier:2015tda}, where the melting temperatures are estimated using a complex potential, indicates that the \upsithree state is expected to melt
essentially at \Tc (where $\Tc=172.5\MeV$ for that study), the \upsitwo state should melt at $T\approx215\MeV$, and the
ground state should survive up to $T\approx460\MeV$. Existing models incorporate several mechanisms leading to the observed bottomonium suppression: screening, thermal decay widths, quarkonium evolution in the high-temperature phase,
regeneration effects, recombination effects, and feed-down contributions \cite{Krouppa:2016jcl, Brambilla:2016wgg, Emerick:2011xu, Du:2017qkv}. The creation of quarkonia from uncorrelated quarks, i.e. recombination, is expected to be negligible for bottomonia compared to expectations for the charmonium family~\cite{Thews:2000rj,Gorenstein:2000ck,Andronic:2007bi,Abelev:2013ila} because the recombination is driven by the number of heavy quark pairs present in a single event, which is much smaller for beauty than for charm. Since the bottom production cross section at 5.02 TeV is of the order of 100--200\unit{$\mu$b}~\cite{Cacciari:2012ny}, this will result in the production of only 2 \bbbar pairs per central nucleon-nucleon collision. By comparison, the charm cross section is of the order of 1\unit{mb} at 200\GeV. Because of the expected small recombination contribution, measurements of \upsi suppression are useful to compare to theoretical calculations of quarkonium in hot nuclear matter and to understand the behavior of quarkonia in high temperature QCD.

Double ratios are useful to quantify the relative modifications of the \upsi excited states.
Theoretically, the uncertainties associated with perturbative QCD calculations (renormalization
and factorization scales, \PQb quark mass, nuclear parton distribution functions) affect the cross sections
in the same way for all \upsi states, and thus cancel in the ratio of excited to ground state yields.
Experimentally, the efficiencies and acceptances cancel almost completely in these double ratios, reducing the measurement uncertainties.

This Letter reports the double ratios
\begin{linenomath}
\begin{equation*}
\frac{(\upsitwo/\upsione)_{\PbPb}}{(\upsitwo/\upsione)_{\pp}}
\quad \text{and} \quad
\frac{(\upsithree/\upsione)_{\PbPb}}{(\upsithree/\upsione)_{\pp}}
\end{equation*}
\end{linenomath}
comparing \pp and \PbPb collisions at a center-of-mass energy per nucleon pair of $\sqrtsnn = 5.02\TeV$, using data collected with the CMS detector during the 2015 LHC run.
The increase in the collision energy and integrated luminosity
allows for a more detailed study compared to the previous measurement at a collision energy of $\sqrtsnn = 2.76\TeV$~\cite{Chatrchyan:2012lxa}.
In particular, we present a more sensitive search for the \upsithree state
in \PbPb collisions and a more accurate measurement of the \upsitwo suppression
in peripheral \PbPb collisions (those with a large impact parameter between the lead ions).
The increase in center-of-mass energy was predicted
to lead to a $\simeq$16\% higher medium temperature~\cite{Krouppa:2016jcl} and to correspondingly stronger suppression effects.

The central feature of the CMS apparatus is a superconducting solenoid of 6\unit{m} internal diameter, providing a magnetic field of 3.8\unit{T}.
Within the solenoid volume are a silicon pixel and strip tracker, a lead tungstate crystal electromagnetic calorimeter, and a brass and scintillator hadron calorimeter, each composed of a barrel and two endcap sections.
Forward calorimeters extend the coverage provided by the barrel and endcap detectors.
Muons are measured in the pseudorapidity range $\abs{\eta} < 2.4$, in gas-ionization detectors embedded in the steel flux-return yoke outside the solenoid, with detection planes made using three technologies: drift tubes, cathode strip chambers, and resistive plate chambers.
Matching muons to tracks measured in the silicon tracker leads to a relative transverse momentum (\pt) resolution between 1 and 2\% for a typical muon in this analysis~\cite{Chatrchyan:2012xi}.
A more detailed description of the CMS detector, together with a definition of the coordinate system used and the relevant kinematic variables, can be found in Ref.~\cite{Chatrchyan:2008zzk}.

For \PbPb collisions, the centrality measurement is based on the sum of transverse energy measured in two hadron forward (HF) calorimeters, which cover the range $2.9 < \abs{\eta} < 5.2$.  In order to select hadronic \PbPb (\pp) collisions, at least three (one) towers with energy deposits above 3\GeV are required in each of the HF calorimeters, both at forward and backward rapidity. A primary vertex reconstructed with at least two tracks is also required. In addition, a filter on the compatibility of the silicon pixel cluster width and the vertex position is applied ~\cite{Chatrchyan:2011sx}. The combined efficiency for this event selection, and the remaining non-hadronic contamination, is $99\pm2\%$. We focus on events where a hard collision is needed in order to produce \upsi\ mesons. Hence, the fraction of such events removed by the minimum-bias trigger requirement is negligible. The event centrality observable corresponds to the fraction of the total inelastic hadronic cross section, starting at 0\% for the most central collisions and evaluated as percentiles of the distribution of the energy deposited in the HF~\cite{Chatrchyan:2011pb}. The average number of nucleons that participate in the interaction for a given centrality class, \Npart, is estimated using a Glauber Monte Carlo (MC) simulation~\cite{Loizides:2014vua}. The Glauber model parameters used for 5.02\TeV \PbPb collisions and a description of the method are given in Ref.~\cite{Khachatryan:2016odn}.

The \upsi mesons are identified via their decay to muons. This analysis uses event samples collected with a dimuon trigger that requires two muons with no explicit single-muon momentum threshold. The same trigger algorithm is used in \pp as well as \PbPb collisions. The algorithm uses information from the muon chambers, which are shielded from the large multiplicities present in \PbPb collisions. Therefore, the performance of the trigger is the same in both collision systems, and across all centralities studied. The trigger sampled an integrated luminosity of 28.0\pbinv in \pp collisions. The \PbPb sample was collected in two ways: by prescaling the dimuon trigger, and by combining the dimuon trigger with an additional selection on 30--100\% centrality collisions. The first setup collected data corresponding to an integrated luminosity of 368\mubinv, and the corresponding dataset is used to derive the centrality-integrated (0--100\%) double ratios and those in the 0--30\% centrality range. For the second setup, the lower rate allowed the sampling of the full integrated luminosity of 464\mubinv. This sample is used to analyze the centrality dependence of the double ratio in the 30--100\% range. We also studied a possible contamination from photo-production processes in the peripheral region and found it to be negligible.

Single muons are selected in the kinematic region $\pt^{\PGm} > 4$\GeVc, $\abs{\eta^{\PGm}}<2.4$, and required to survive standard quality selection criteria~\cite{Chatrchyan:2012xi}. The reconstruction algorithm was adapted to account for the high track multiplicity in a \PbPb event, using a combination of regional and iterative tracking algorithms~\cite{Chatrchyan:2014fea}.
The muon momentum is derived from the fit obtained with a Kalman filter algorithm~\cite{Chatrchyan:2012xi}
applied to the tracker hits and provides an $\upsi$ mass resolution of around 1\% in both \pp and \PbPb. When forming a muon pair, the two reconstructed muon candidates are required to match the dimuon trigger and to originate from a common vertex with a $\chi^2$ probability larger than 1\%. The \upsi\ transverse momentum and rapidity ranges studied in this analysis are $\pt<30$\GeVc and $\abs{y}<2.4$. The \upsi ratios are not affected by the small number of additional collision vertices (pileup) present in the \pp and \PbPb\ samples.

\begin{figure}[htb]
\centering
  \includegraphics[width=0.48\textwidth]{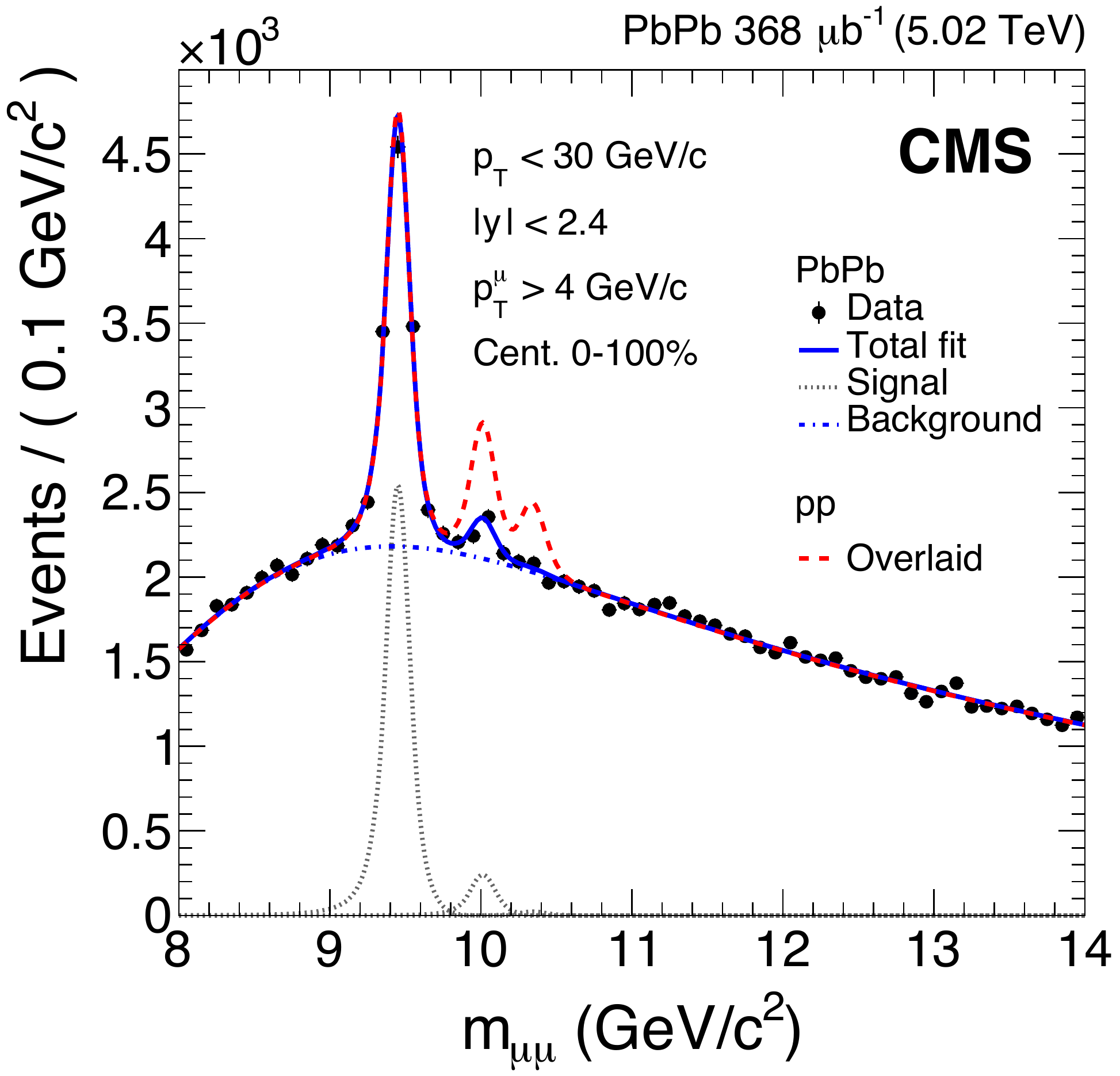}
  \caption{Measured dimuon invariant mass distribution in \PbPb data.
  The total fit (solid blue line) and the background component (dot-dashed blue line) are also shown, as are
  the individual \upsione, \upsitwo, and \upsithree signal shapes (dotted gray lines).
  The dashed red line represents the \pp signal shape added to the \PbPb background and normalized to the \upsione mass peak in \PbPb.
  \label{fig:InvMass}}

\end{figure}

Figure~\ref{fig:InvMass} shows the invariant mass distributions of opposite-charge muon pairs
for centrality-integrated \PbPb collisions. The double ratios are computed from
the signal yields obtained independently from unbinned maximum likelihood fits to the \pp and \PbPb spectra. The analysis of the \upsitwo double ratio is performed in three $\pt$ bins, two $\abs{y}$ bins, and nine centrality bins, while the \upsithree double ratio is studied in four centrality bins. As a cross-check, simultaneous fits of the two dimuon invariant mass distributions, where the double ratios are directly extracted, were also performed. The two procedures give consistent results.

The shape of each \upsi state is modeled with the sum of two Crystal Ball functions~\cite{Oreglia:1980cs}, with parameters fixed from MC simulation studies. The mass parameter of the \upsione resonance is left free to account for possible shifts in the momentum scale of the reconstructed tracks, and is found to be consistent between  \pp and \PbPb data. The masses of the excited states are fixed to the \upsione mass scaled by the ratio of the world average mass values~\cite{Olive:2016xmw}. The systematic uncertainty in the double ratio from the choice of signal model is evaluated by testing two fit variations. One uses the same function, but allowing all previously fixed parameters to float one by one and propagating as systematic uncertainty the maximum observed deviations from the double ratios obtained with the nominal signal model. The second fit variation uses a sum of a Crystal Ball function and Gaussian function as an alternative fit model. The total uncertainties related to the signal model are determined by summing in quadrature the two systematic components, and are in the ranges 1--10\% and 9--15\% for the \upsitwo and \upsithree double ratios, respectively.

The background is modeled with an error function multiplied by an exponential function as in Ref.~\cite{Chatrchyan:2012lxa}, a parameterization selected, in each analysis bin, through a log-likelihood ratio test comparing several functional forms, while fixing the signal parameters. For the two highest \pt bins in this analysis, using an exponential without the error function provides the best fit.
Possible deviations in the results when choosing an alternative background model,
in the form of a 4th-order polynomial, are studied using pseudo-experiments. For this purpose, the nominal background and signal models are used to generate pseudo invariant mass distributions in each bin of the analysis. These distributions are then fit with the nominal model as well as using the alternative background model. The average resulting differences between nominal and alternative fit model are found to be in the 2-15\% range for the \upsitwo and \upsithree double ratios, respectively. The signal and background model uncertainties are the dominant sources of systematic uncertainty in this analysis.

Possible effects of non-cancellation of reconstruction, trigger, and muon identification efficiencies in the double ratios are studied by comparing
the results of simulations using \PYTHIA 8.209~\cite{Sjostrand:2007gs} tune CUETP8M1 (for the low-occupancy \pp environment) with those obtained using \PYTHIA8 embedded in \textsc{hydjet} 1.9~\cite{Lokhtin:2005px} (for the high-occupancy \PbPb data).
The \upsi transverse momentum distributions in the MC samples are reweighted to match the signal \pt spectra seen in data, since the reconstruction efficiency depends on $\pt$. The rapidity distributions in simulation are consistent with those in data, hence no reweighting is applied as a function of $y$. The maximum deviation from unity of the double ratio of efficiencies, among all the analysis bins, was found to be 1.4\%, a value taken as a systematic uncertainty.

Acceptance corrections are not applied because they are expected to cancel in the \PbPb over \pp ratio for each state.
If, however, the \upsi meson acceptances were different in \pp and \PbPb because of physical effects, such as a change in polarization or strong kinematical differences from \pp to \PbPb collisions within an analysis bin, these would not cancel in the double ratio.
The hypothesis that such potential effects can be neglected is supported by the absence of significant changes of the \upsin polarizations in \pp collisions as a function of event activity~\cite{Khachatryan:2016vxr}.
Moreover, when studying the $\pt$ and $\abs{y}$ distributions in the \pp and \PbPb data samples, it is observed that they have similar shapes. As in previous analyses~\cite{Chatrchyan:2011pe,Chatrchyan:2012lxa,Chatrchyan:2013nza,Khachatryan:2014bva,Sirunyan:2016znt}, possible differences in \PbPb and \pp acceptances due to physical effects are not considered as systematic uncertainties.

\begin{figure}[htb]
\centering
  \includegraphics[width=0.48\textwidth]{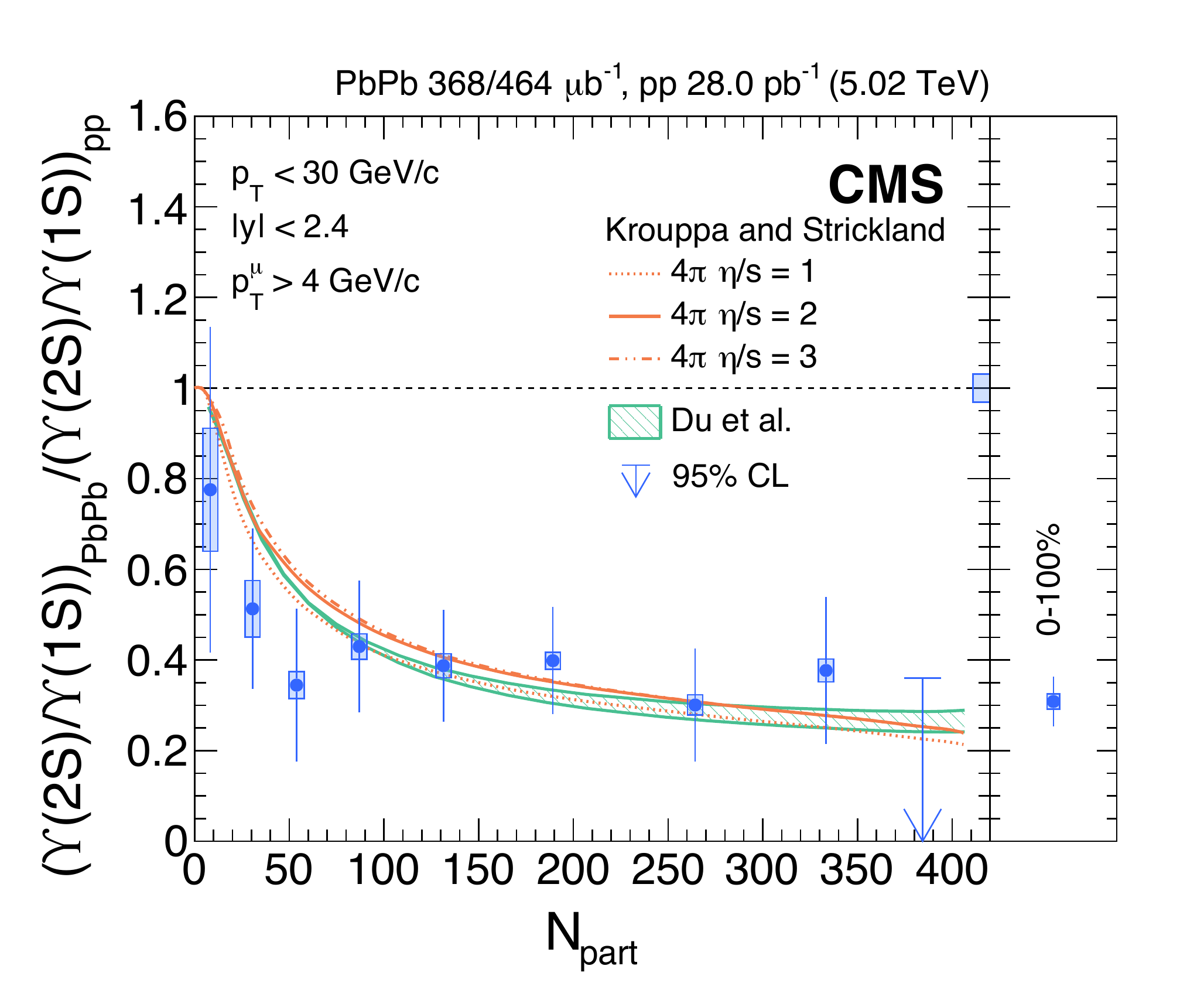}
  \caption{Double ratio of the \upsitwo as a function of centrality. The
	  centrality-integrated value is shown in the right panel.
	  The error bars represent the statistical uncertainty in the \PbPb data while the boxes represent the systematic uncertainty due to signal and background variations. The global systematic uncertainty is represented as a grey box drawn around the line at unity.
	  Calculations by Krouppa and Strickland (orange curves~\cite{Krouppa:2016jcl}) and by Du et al. (green hatched band~\cite{Emerick:2011xu,Du:2017qkv}) are also shown.
  \label{fig:DR21}}

\end{figure}

Figure~\ref{fig:DR21} shows the \upsitwo double ratio as a function of \Npart.
The box drawn around the line at unity represents the global uncertainty, that applies to all measurements, including the centrality-integrated datum point. It amounts to 3.1\%, and includes the systematic and statistical uncertainties from the \pp single ratio, as well as the uncertainty due to possible non-cancellation of reconstruction, trigger, and muon-identification efficiencies. A large relative suppression of the \upsitwo state compared to the \upsione state in \PbPb collisions with respect to the \pp data is observed.
The centrality-integrated \upsitwo double ratio is
$0.308 \pm 0.055\stat \pm  0.019\syst$, where the systematic uncertainty reflects the signal and background variations in \PbPb and \pp data, as well as the uncertainty on the combined detection efficiency.
In the most peripheral bin (70--100\%), the double ratio is consistent with unity.
In the most central bin (0--5\%), the \upsitwo signal is consistent with zero within one standard deviation of the statistical uncertainty.
Therefore, a 95\% confidence level (\CL) interval is derived for this
centrality bin, obtained using the Feldman--Cousins method~\cite{Feldman:1997qc}. The relative  \upsitwo suppression is similar at 5.02 and 2.76\TeV~\cite{Chatrchyan:2012lxa}. The results presented here have an improvement in the statistical precision compared to the previous measurement by almost a factor of 2.

Predictions of \upsi suppression from Krouppa and Strickland~\cite{Krouppa:2016jcl}, incorporating color-screening effects on the bottomonium family and reflecting feed-down contributions from decays of heavy quarkonia, are in overall agreement with the \upsitwo double ratio results presented in Fig.~\ref{fig:DR21}. In this model, the dynamical evolution is treated using anisotropic hydrodynamics, where the relevant initial conditions are changed by varying the viscosity to entropy ratio, $\eta/s$, and the initial momentum-space anisotropy. In order to maintain agreement with charged multiplicity and elliptic flow measurements, the initial temperature is then uniquely determined as well. The temperatures reported in this model are in the range $T={641, 631, 629}$\MeV corresponding to $4\pi\eta/s={1,2,3}$, respectively. Another theoretical curve from Du \etal~\cite{Du:2017qkv}, based on a kinetic-rate equation approach first presented in Ref.~\cite{Emerick:2011xu} and containing a small component of regenerated bottomonia, shows a similar level of agreement with the data.
In this model, the absence of a regeneration component would lead to almost complete suppression of the \upsitwo, \ie, a double ratio of zero for the centrality
range $\Npart > 250$. Such a scenario is ruled out by our data.

\begin{figure}[htb]
\centering
  \includegraphics[width=0.48\textwidth]{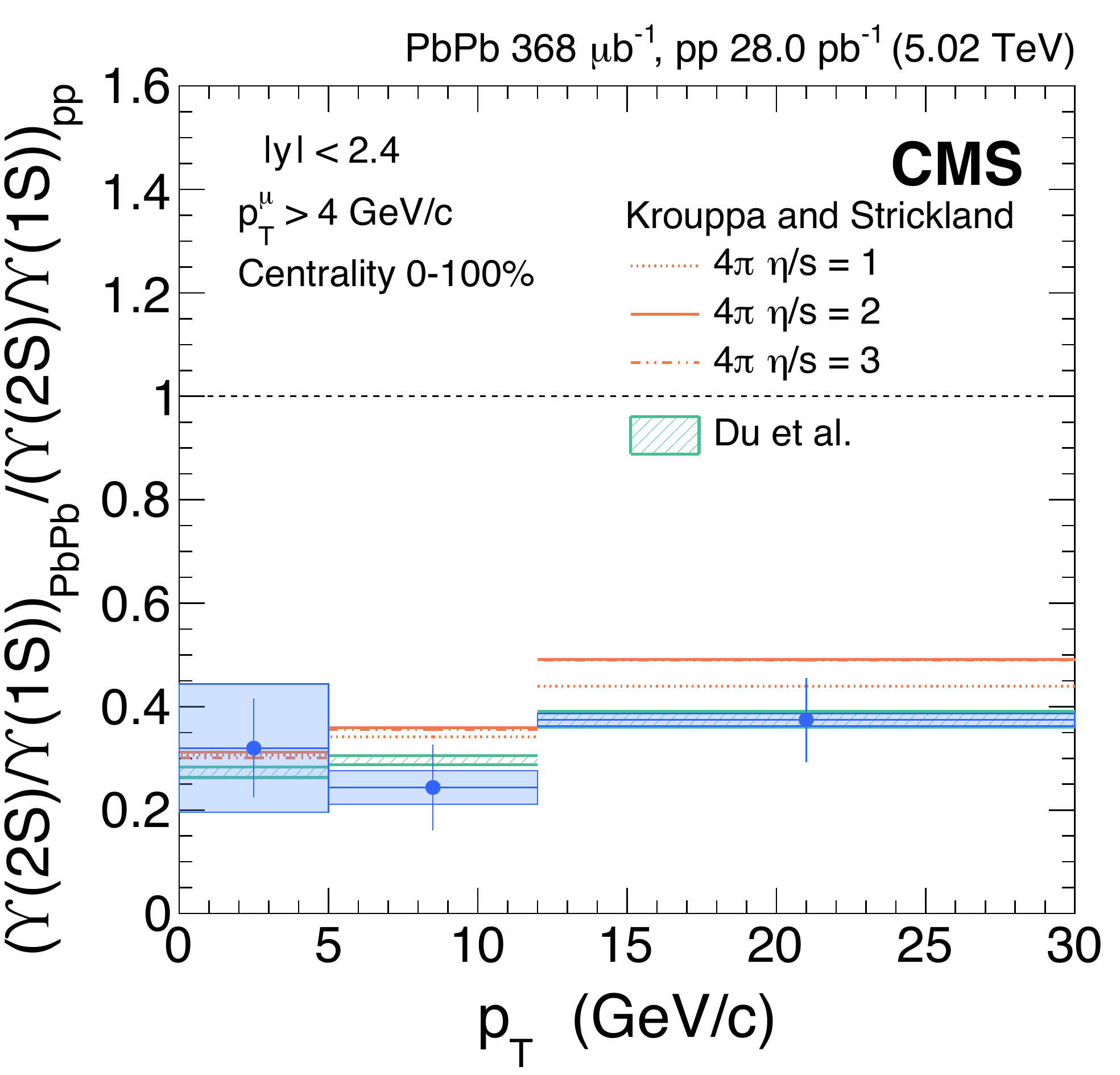}
  \includegraphics[width=0.48\textwidth]{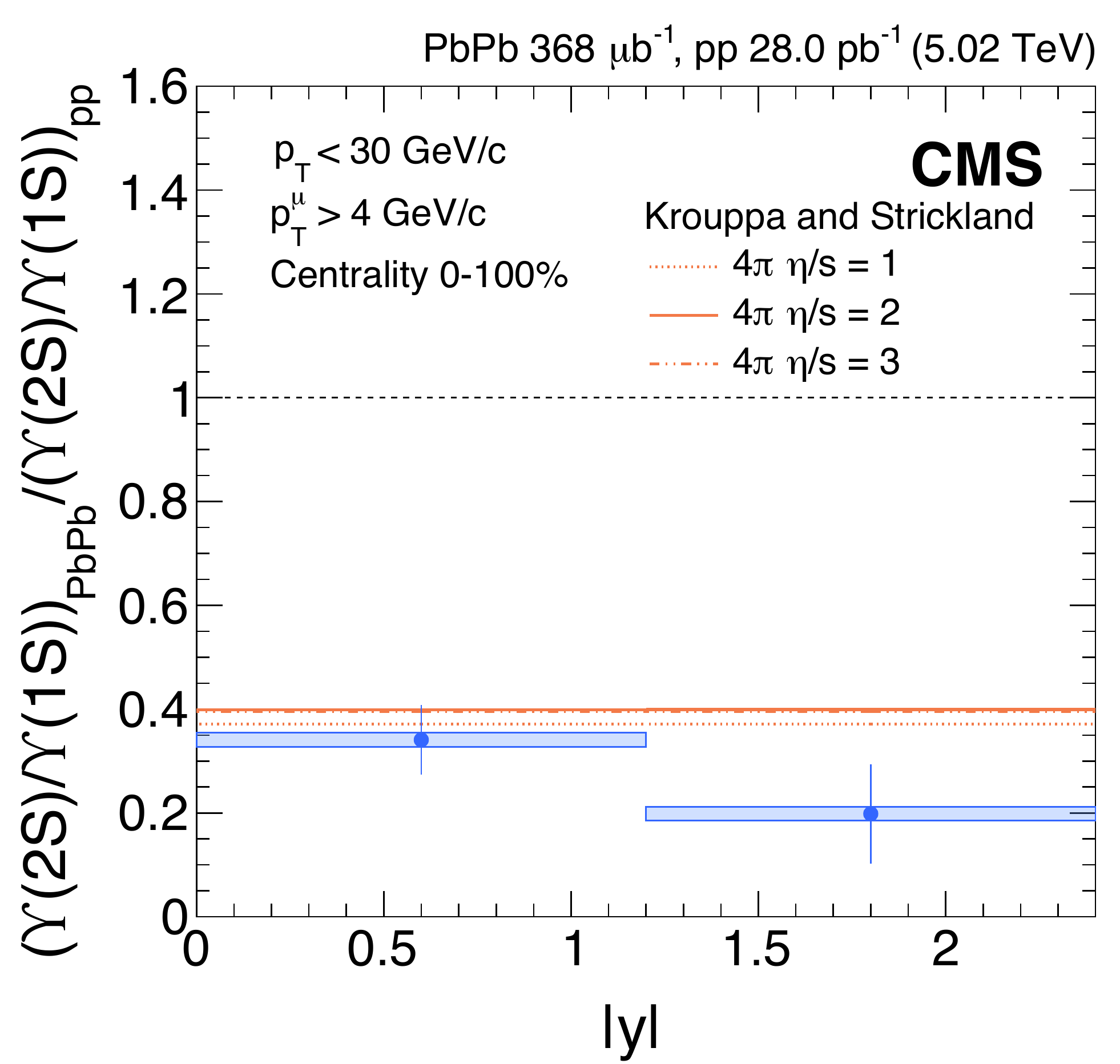}
  \caption{Double ratio of the \upsitwo  as functions of $\pt$ (\cmsLeft) and $\abs{y}$ (\cmsRight).
	  The error bars depict the statistical uncertainty while the boxes represent the systematic uncertainties in the signal and background models as well as the combined detection efficiency.  Calculations by Krouppa and Strickland (orange curves~\cite{Krouppa:2016jcl}) and by Du et al. (green hatched band~\cite{Emerick:2011xu, Du:2017qkv}) are also shown.
  \label{fig:DR21ptRap}}
\end{figure}

Figure~\ref{fig:DR21ptRap} shows the \upsitwo double ratio as a function of $\pt$ and $\abs{y}$.
A large relative \upsitwo suppression is observed throughout the kinematic range studied, with no significant variations with $\pt$ or $\abs{y}$. Predictions of $\upsi$ suppression as functions of $\pt$~\cite{Krouppa:2016jcl,Emerick:2011xu} and $\abs{y}$~\cite{Krouppa:2016jcl} are in overall agreement with the data.

\begin{figure}[htb]
\centering
  \includegraphics[width=0.50\textwidth]{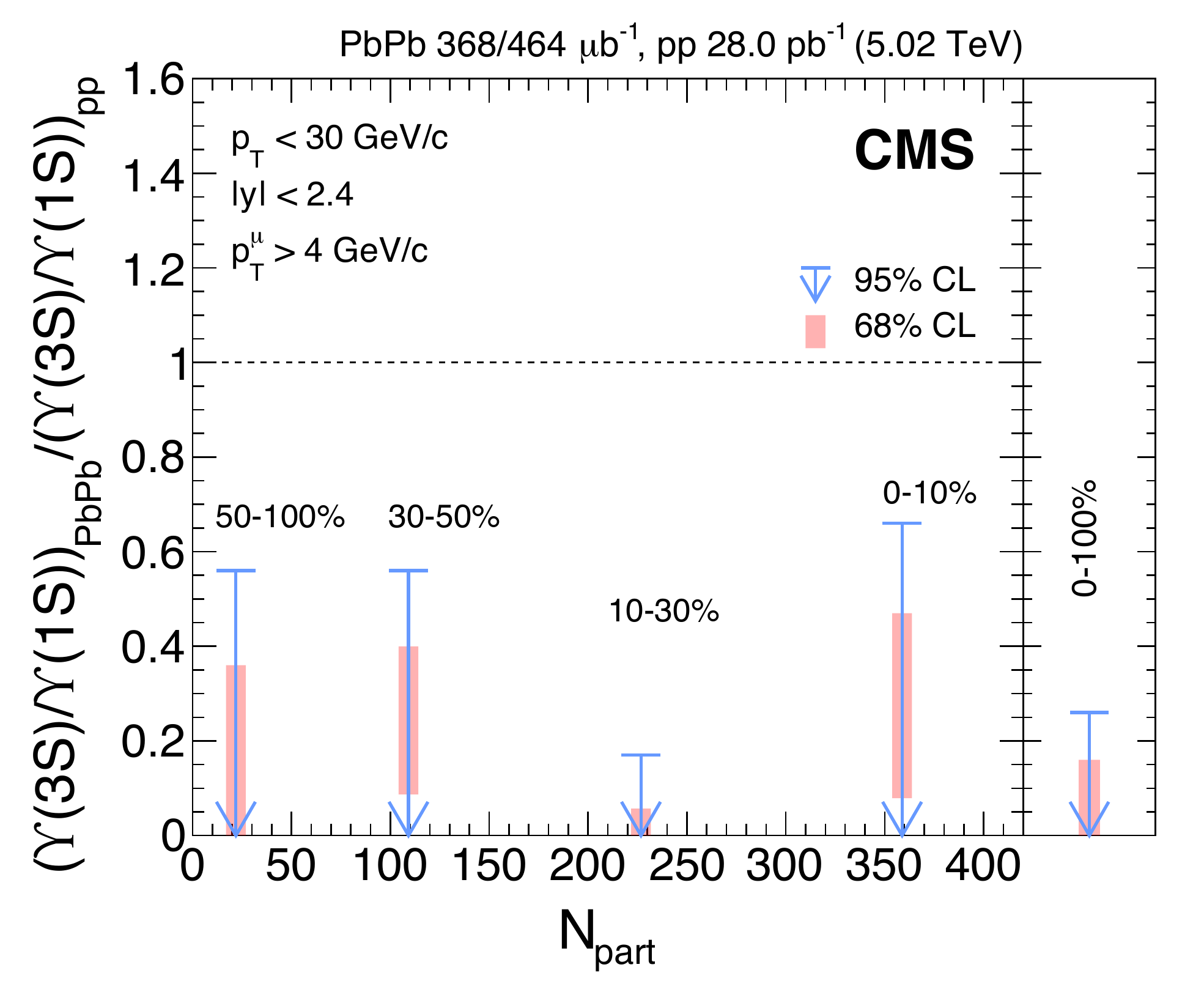}
  \caption{Confidence intervals at 95\% \CL (blue arrows) and 68\% \CL (red boxes) of the \upsithree double ratio as a function of centrality. The centrality-integrated limit is shown in the right panel.
  \label{fig:DR31}}
\end{figure}

For the \upsithree, as seen in Fig.~\ref{fig:InvMass}, the signal yield in the \PbPb data is consistent with zero in the centrality-integrated sample.
Figure~\ref{fig:DR31} shows the extracted \upsithree double-ratio confidence intervals,
at 95\% and 68\% \CL. In all four centrality bins, the \upsithree double ratio is significantly below unity,
showing that the \upsithree state is strongly suppressed relative to the \upsione state, even in the most peripheral (50--100\%) \PbPb collisions probed in this analysis.
The centrality-integrated \upsithree double ratio is smaller than 0.26 at 95\% \CLp.
We excluded the possibility that the stringent limit in the 10--30\% centrality bin is due to a large downward fluctuation in the background by studying
the invariant mass region of the \upsithree\ in each centrality bin.
We also calculated upper limits
under the assumption that the observed counts are equal to the expected background and found an upper limit that increases only slightly
to the range 0.2--0.3 for the 10--30\% bin.

In summary, the \upsitwo and \upsithree double ratios have been measured at 5.02\TeV, using \pp and \PbPb data samples significantly larger than those used in the corresponding 2.76\TeV measurements.
The centrality-integrated double ratios are
  $0.308 \pm 0.055\stat \pm  0.019\syst $
 for the \upsitwo and
 $<$0.26 at 95\% \CL for the \upsithree.
The large relative suppression of the \upsitwo does not show significant variations with $\pt$ or $\abs{y}$ within the
explored phase space window of $\pt< 30\GeVc$ and $\abs{y}<2.4$. The \upsitwo double ratio is compatible with unity in the most peripheral collisions (70--100\%) and with zero in the most central ones (0--5\%), but a flat centrality dependence is not excluded, given the current uncertainties. The 95\% \CL intervals for the \upsithree double ratio exclude unity in the four centrality bins of this analysis, including the most peripheral collisions (50--100\%).

\begin{acknowledgments}
We congratulate our colleagues in the CERN accelerator departments for the excellent performance of the LHC and thank the technical and administrative staffs at CERN and at other CMS institutes for their contributions to the success of the CMS effort. In addition, we gratefully acknowledge the computing centers and personnel of the Worldwide LHC Computing Grid for delivering so effectively the computing infrastructure essential to our analyses. Finally, we acknowledge the enduring support for the construction and operation of the LHC and the CMS detector provided by the following funding agencies: BMWFW and FWF (Austria); FNRS and FWO (Belgium); CNPq, CAPES, FAPERJ, and FAPESP (Brazil); MES (Bulgaria); CERN; CAS, MoST, and NSFC (China); COLCIENCIAS (Colombia); MSES and CSF (Croatia); RPF (Cyprus); SENESCYT (Ecuador); MoER, ERC IUT, and ERDF (Estonia); Academy of Finland, MEC, and HIP (Finland); CEA and CNRS/IN2P3 (France); BMBF, DFG, and HGF (Germany); GSRT (Greece); OTKA and NIH (Hungary); DAE and DST (India); IPM (Iran); SFI (Ireland); INFN (Italy); MSIP and NRF (Republic of Korea); LAS (Lithuania); MOE and UM (Malaysia); BUAP, CINVESTAV, CONACYT, LNS, SEP, and UASLP-FAI (Mexico); MBIE (New Zealand); PAEC (Pakistan); MSHE and NSC (Poland); FCT (Portugal); JINR (Dubna); MON, RosAtom, RAS, RFBR and RAEP (Russia); MESTD (Serbia); SEIDI, CPAN, PCTI and FEDER (Spain); Swiss Funding Agencies (Switzerland); MST (Taipei); ThEPCenter, IPST, STAR, and NSTDA (Thailand); TUBITAK and TAEK (Turkey); NASU and SFFR (Ukraine); STFC (United Kingdom); DOE and NSF (USA).
\end{acknowledgments}
\bibliography{auto_generated}

\cleardoublepage \appendix\section{The CMS Collaboration \label{app:collab}}\begin{sloppypar}\hyphenpenalty=5000\widowpenalty=500\clubpenalty=5000\textbf{Yerevan Physics Institute,  Yerevan,  Armenia}\\*[0pt]
A.M.~Sirunyan, A.~Tumasyan
\vskip\cmsinstskip
\textbf{Institut f\"{u}r Hochenergiephysik,  Wien,  Austria}\\*[0pt]
W.~Adam, E.~Asilar, T.~Bergauer, J.~Brandstetter, E.~Brondolin, M.~Dragicevic, J.~Er\"{o}, M.~Flechl, M.~Friedl, R.~Fr\"{u}hwirth\cmsAuthorMark{1}, V.M.~Ghete, C.~Hartl, N.~H\"{o}rmann, J.~Hrubec, M.~Jeitler\cmsAuthorMark{1}, A.~K\"{o}nig, I.~Kr\"{a}tschmer, D.~Liko, T.~Matsushita, I.~Mikulec, D.~Rabady, N.~Rad, B.~Rahbaran, H.~Rohringer, J.~Schieck\cmsAuthorMark{1}, J.~Strauss, W.~Waltenberger, C.-E.~Wulz\cmsAuthorMark{1}
% ...
\vskip\cmsinstskip
% ...
1:~~Also at Vienna University of Technology, Vienna, Austria\\
% ...

\end{sloppypar}
\end{document}